\documentstyle[osa,manuscript]{revtex}

\begin{document}
\title{Magnetic remanence of Josephson junction arrays}
\author{W. A. C. Passos, F. M. Araujo-Moreira and W. A. Ortiz}
\address{Grupo de Supercondutividade e Magnetismo, Departamento de F\'{i}sica\\
Universidade Federal de S\~{a}o Carlos, Caixa Postal 676 - 13565-905 S\~{a}o%
\\
Carlos, SP, Brazil}
\maketitle

\begin{abstract}
In this work we study the magnetic remanence exhibited by Josephson junction
arrays in response to an excitation with an AC magnetic field. The effect,
predicted by numerical simulations to occur in a range of temperatures, is
clearly seen in our tridimensional disordered arrays. We also discuss the
influence of the critical current distribution on the temperature interval
within which the array develops a magnetic remanence. This effect can be
used to determine the critical current distribution of an array.


\end{abstract}

%
%
%

\bigskip

In 1997, Araujo-Moreira and coworkers\cite{araujo} conducted a systematic
investigation on the AC magnetic susceptibility ($\chi _{AC}$) of Josephson
junction arrays (JJAs) as a function of temperature (T) and excitation field
(h). Relying on simulations, they have explained the dynamic reentrance (DR)
displayed by $\chi _{AC}$(T). The authors have also demonstrated that JJAs
may exhibit a paramagnetic response, offering thus a plausible explanation
for the Wohlleben effect (WE) - also called Paramagnetic Meissner effect -
which has been unsystematically observed in several samples of high (HTS)
and low (LTS) temperature superconductors. Based on this interpretation, the
WE behavior would be due to one or more JJAs being formed on those samples,
depending on the strict conditions under which the specimens were prepared.

In a recent work\cite{ortiz}, we have confirmed that WE is an inherent
property of JJAs. In another front line\cite{passos}, we have also
demonstrated that tridimensional (3D) JJAs can be produced in a controlled
manner. The 3D-JJAs exhibit both WE and DR. In their simulations using a
single plaquette, Araujo-Moreira et alli\cite{araujo} have shown that this
simple model accounts for the magnetic behavior of a JJA excited by AC
magnetic fields. The magnetic induction versus applied field curve (B versus
H) predicts that the array may exhibit a remanent moment upon excitation by
a magnetic field.

This contribution reports a systematic study of the magnetic moment
displayed by a JJA after excitation by an AC field of variable amplitude.
The arrays were fabricated from LTS granular superconductors. The
experimental results confirm those anticipated by simulations, including the
fact that the remanence only appears for a limited range of temperatures.

We have selected a 3D-JJA from a batch of LTS granular samples fabricated
following a standard procedure described elsewhere\cite{passos}. In short,
niobium powder is separated according to grain size, using a set of special
sieves, with mesh gauges ranging from 38 to 44 $\mu $m. The powder is then
uniaxially pressed in a mold to form a cylindrical pellet of 2.5 mm radius
by 2.0 mm height. This pellet is a tridimensional disordered JJA (3D-DJJA)
in which the junctions are weakly-coupled grains, i.e., weak-links formed by
a sandwich between Nb grains and a Nb-oxide layer originally present on the
grain surface. As a consequence of the uniaxial pressure, samples produced
in this way are anisotropic. This anisotropy can be enhanced by applying
higher uniaxial pressures or, in contrast, it can be decreased through
submission of the pellet to an isostatic pressure. In fact, the critical
currents (J$_{c}$) of the weak-links laying in planes perpendicular to the
pressure axis are much less dispersed than those along that axis, as
detected by measurements of the magnetic AC susceptibility. Further
treatment under isostatic pressures tends to homogenize the critical
currents and eliminate the anisotropy\cite{passos}.

Before pelletizing, powder quality is monitored through several techniques
as X-ray diffraction (XRD), scanning electron microscopy (SEM) imaging and
magnetic measurements ($\chi _{AC}$ and M\cite{moment} versus T). Proper
characterization of the grains certify that we have good-quality Nb without
contaminants, probably with some interstitial gases, as the critical
temperature (T$_{c}$) is 8.9 K. Also, size distribution of the grains is
within the mesh range of the sieves, adequate for the purposes of the study.

To carry out the present work, we have selected an anisotropic 3D-JJA, so
that we could not only study the magnetic remanence of the system, but also
emphasize the role of J$_{c}$ dispersion and intensity on the observed
effect. The sample exhibits all characteristic features of a genuine 3D-JJA.
Two of these peculiarities are shown in Figure 1: the main picture is a
low-field measurement of the reentrant magnetization (WE), for H = 2 Oe. The
inset displays a Fraunhofer pattern for the real part of $\chi _{AC}$ which,
as discussed in Ref. 5, is an indirect determination of the J$_{c}$.

The above mentioned prediction that JJAs may develop a magnetic remanence
can be appreciated in Figure 2. It is a sketch of simulation results\cite
{araujo} of the B versus H curves (B = H + 4$\pi $M), at temperatures T$_{1}$
\mbox{$<$}%
\ T$_{2}$ 
\mbox{$<$}%
\ T$_{3}$ 
\mbox{$<$}%
\ T$_{c}$, for a plaquette with four identical JJs. For T = T$_{2}$ one can
see that, if the applied field H reaches a value larger than H$_{0}$ and
then is turned off, the system will retain a magnetic moment for H = 0. On
the contrary, there will be no remanence at temperatures T$_{1}$ and T$_{3}$%
, irrespective of the maximum value attained by H. It must be also noticed
that the remanence predicted for T = T$_{2}$ does not depend on the time
profile of the applied field, which could have, for example, a triangular
shape (up and down) or be a sinusoidal excursion around zero, as long as its
magnitude reaches the threshold value H$_{0}$ before returning to zero.

To verify the validity of these predictions, we performed a series of
experiments, based on the following steps:

i. the sample is submitted to an AC field (h) consisting of a train of
sinusoidal pulses, after what h is kept null;

ii. with h = 0, the magnetic moment of the sample is measured.

Steps (i) and (ii) are the core of two experimental routines: the field scan
routine (FS), for which the field is changed at a fixed temperature, and the
temperature scan routine (TS), when one changes T at a fixed value of h.

Measurements following routines FS and TS were performed using a Quantum
Design MPMS-5T SQUID magnetometer. Both routines were extensively explored,
furnishing valuable results for the purposes of this work. In this short
paper, however, we will present in more detail results derived from the FS
routine, employed at several temperatures. As shown below, one can easily
recognize the existing connections between these results and the prediction
of a magnetized state, as pictured in Figure 2. Remaining parts of this
study, including many other aspects of the problem, will be published
elsewhere.

Using the FS routine we measured the remanent magnetization (M$_{r}$) as a
function of the excitation field. For an ordinary superconductor of any
kind, from a single crystal to a totally disordered granular sample, the
only possibility of a remanence after the application of the AC field would
be a residual magnetization due to flux eventually pinned inside the
specimen. This contribution, however, is expected to be small and
practically independent of the excitation field. We have verified the above
characteristics measuring M$_{r}$(h, T) for a variety of samples. In
particular, the powder used to fabricate our arrays have the typical
response of ordinary superconductors, so that the effects described below
are entirely due to the formation of the 3D-DJJA.

The inset of Figure 3 shows three examples of measurements performed using
the FS routine. Notice that all curves originate at h = 1 mOe with the same
value of M$_{r}$, but the evolution of each one depends strongly on T. The
response is flat and reversible for low temperatures, but hysteretic in an
interval of higher temperatures. For the sake of clarity, a logarithmic
scale was used for the field h. Curves in the inset are ''as measured'',
i.e., without subtraction of the powder response. The main frame of the
figure displays curves for different values of the excitation field,
constructed using the ascending branch (field up) of the FS routines. For
each value of the field, M$_{r}$ is the remanence of the sample from which
we subtracted the signal of the powder (not shown). It is evident that, as
predicted, M$_{r}$ is confined to a window of temperature, being larger and
more pronounced as the exciting field increases. There is some activity
present at lower temperatures, most possibly related to the unscreened DC
field, of the order of 30 mOe during the experiments. In fact, the
plaquettes probed by the experiment (families of 2D-JJAs perpendicular to
the exciting field), are also penetrated by this additional flux, which
induces screening currents that contribute to the magnetization of the whole
sample. This contribution might be either positive or negative, depending on
a delicate compromise involving many parameters of the system\cite{araujo}$%
^{,}$\cite{passos}$^{,}$\cite{barbara}.

Another effective way of eliminating spurious contributions arising from
pinning - either inside the grains or by the plaquettes - is to take the
difference between measurements on the ascending and descending branches of
the FS routine curves. The main illustration in Figure 4 shows the results
of this procedure in two situations: the squares represent measurements
taken with the field parallel (h$_{//}$) to the pressure axis, whereas the
circles were taken with h perpendicular (h$_{\perp }$) to that axis. Notice
that each curve was normalized to its larger value, based on the ''as
measured'' data presented in the inset of the figure. Not surprisingly, the h%
$_{//}$ curve resembles those shown in Figure 3 (all taken at the same
relative orientation), including the non-zero constant response for lower
temperatures. Most remarkably, however, is the fact that both curves
coincide exactly above the maxima, being rather different in the temperature
interval between 5 K to 7.5 K. This distinction is deeply related to the
anisotropy of the uniaxially pressed sample, and disappears for isotropic
specimens pressed isostatically. As a matter of fact, we have observed for
this particular sample, using SEM imaging, that the pressure used was enough
to deform, along the pressure axis, a significant portion of the grains.
This caused the weak-links to become too much strong, practically soldering
some neighbor grains and destroying part of the plaquettes laying in planes
parallel to the pressure axis. On the other direction, however, the grains
were not deformed and a much larger number of weak-links was preserved.
Thus, one should expect a much more intense peak for the h$_{//}$ curve,
since the intensity reflects the effective number of active plaquettes. The
inset shows the absolute values, giving a peak ratio of the order of \ 4.5.

On the other hand, the window of temperatures for which a remanence is
expected (represented by T$_{2}$ in Figure 3) is dependent on the J$_{c}$%
\cite{araujo}. Thus, the inhomogeneous soldering of the grains along the
pressure axis should also have some influence, as it causes the critical
currents of the weak-links to be much more dispersed than in the
perpendicular plane. As we see from Figure 4, the range of temperatures for
which a remanence exists is much larger for the planes of arrays parallel to
the pressure axis, a direct consequence of the critical current dispersion.
We have, therefore, another important feature of remanence studied here: its
measurement, when properly calibrated, might be taken as a figure of merit
to qualify a JJA in terms of the J$_{c}$ distribution of its constituent
elements.

In conclusion, we have measured the predicted magnetic remanence of JJAs,
using a 3D-DJJA fabricated from granular Nb. The remanence occurs in a
limited interval of temperatures, which extent depends on the excitation
field. The profile of M$_{r}$ is sensitive to the critical current
dispersion, as revealed by experiments with our anisotropic sample, thus
constituting a prospective tool to determine the J$_{c}$ distribution.

We acknowledge financial support from Brazilian agencies CAPES and FAPESP.
We also thank FAENQUIL-Lorena for providing Nb powder.

\bigskip %

\begin{figure}[tbp]
\caption{Magnetization curve of 3D-JJA showing WE. The inset shows a
Fraunhofer pattern of this sample, taken by AC susceptibility measurements.}
\end{figure}

\begin{figure}[tbp]
\caption{Numerical simulation of B versus H for a plaquette with four JJs
[Ref. 1]. The simulation predicts the appearance of a remanence after
application of a field H $\leq $ H$_{0}$ in a range of temperatures (as T = T%
$_{2}$).}
\end{figure}

\begin{figure}[tbp]
\caption{The main illustration shows M$_{r}$ - M$_{powder}$ as a function of
temperature for different magnitudes of the excitation field. The inset
presents the ''as measured'' remanence as a function of h for different
temperatures.}
\end{figure}

\begin{figure}[tbp]
\caption{Sample anisotropy revealed by measurements of the remanence versus
temperature for different orientations of the excitation field. Main graph:
data normalized to peak values. Inset: ''as measured'' data.}
\end{figure}


\begin{references}
\bibitem{moment}  Throughout the paper, M is the volume magnetization, i.e.,
the magnetic moment of the samples.
\end{references}
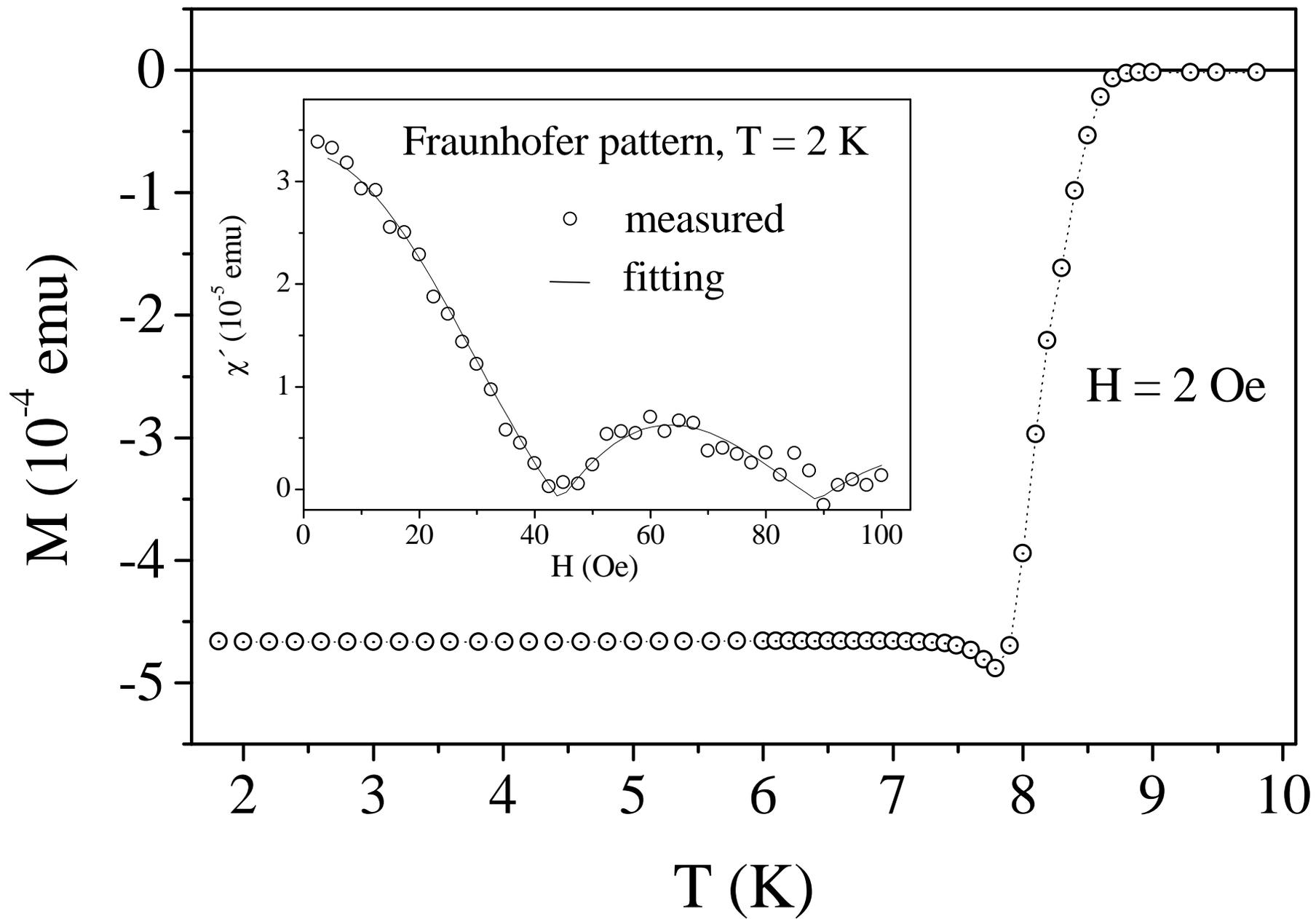
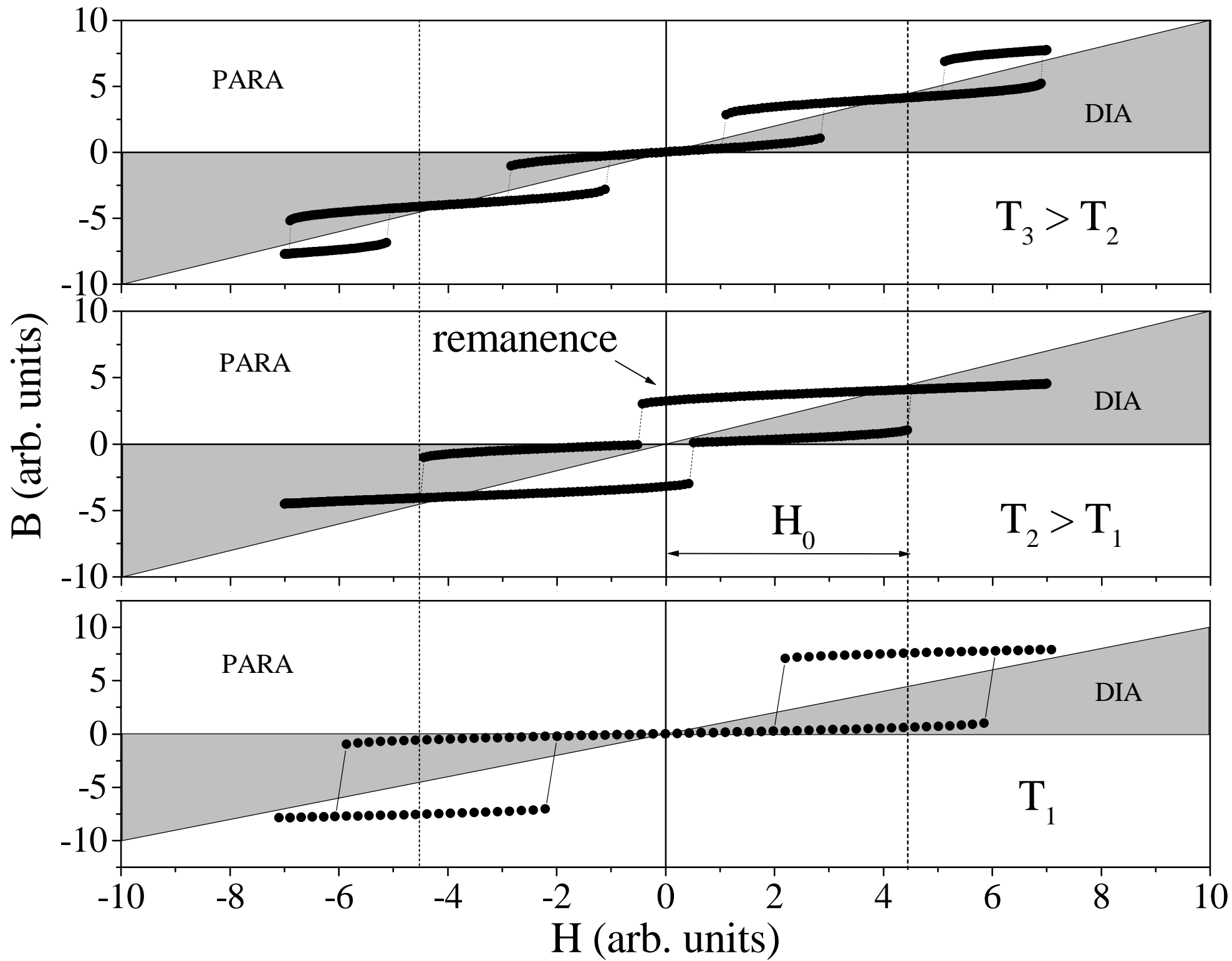
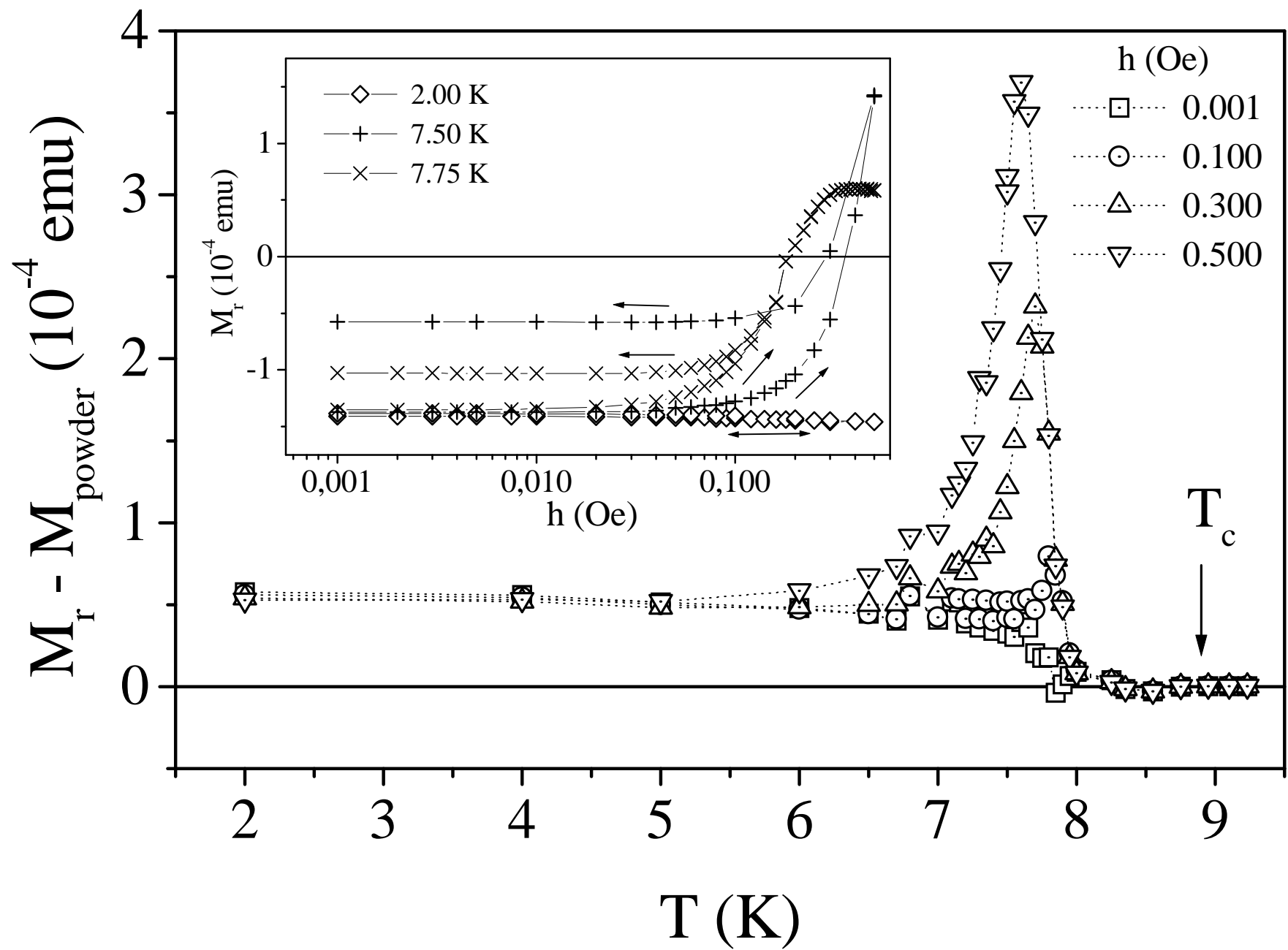
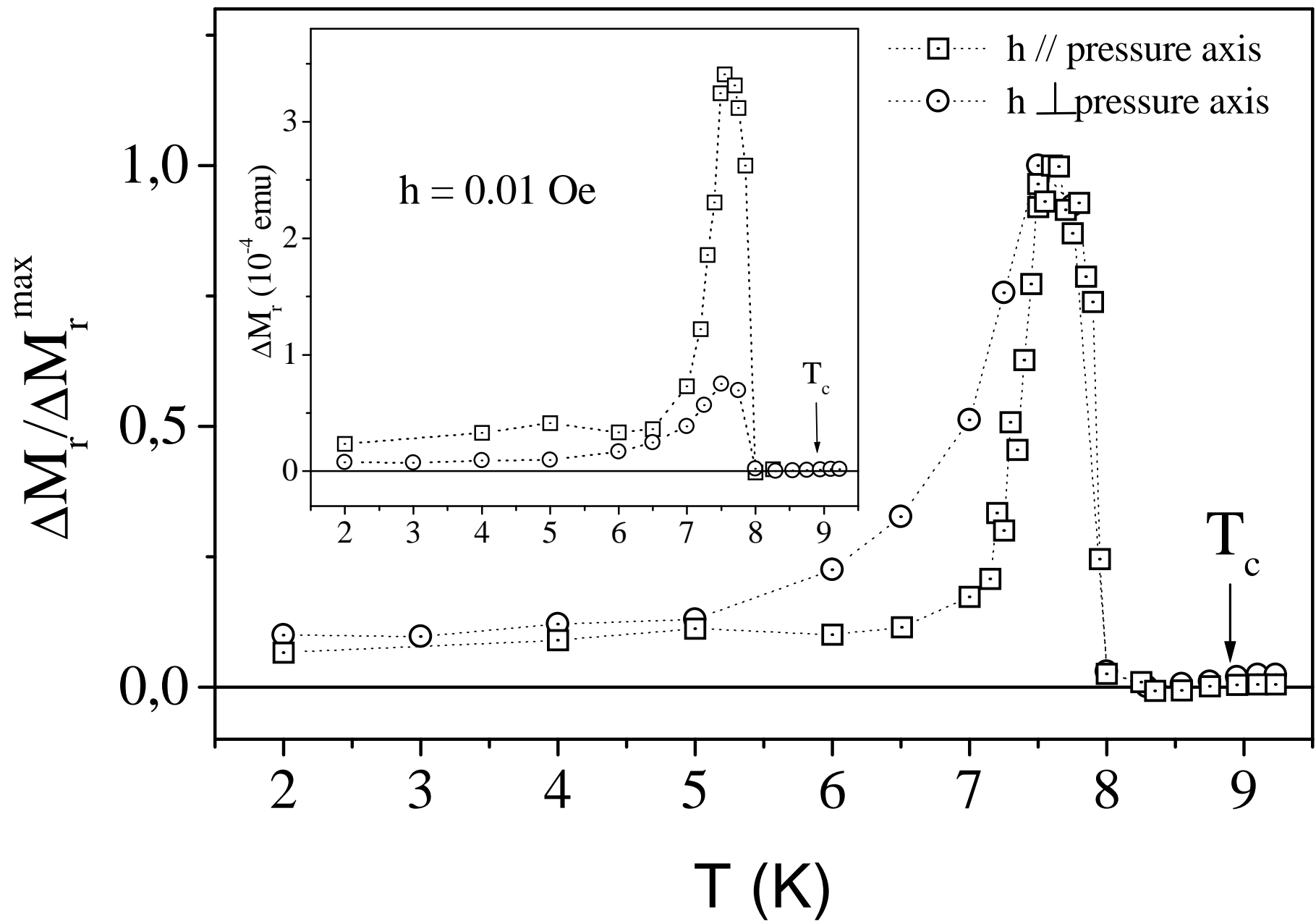
\end{document}